\documentclass[12pt]{article}
\begin{document}

\title{\textbf{Generalized coherent and intelligent states for exact solvable
quantum systems}}
\author{A. H. El Kinani$^{\dagger}$ and M. Daoud$^{\ddagger }$\vspace{0.5cm}
\\ $^{\dagger }$L. P. T, Physics Department, Faculty of Sciences,
\\ University Mohammed V, P.O.Box 1014\\ Rabat,
Morocco.\vspace{1cm}\\ $^{\ddagger }$L. P. M. C, Physics
Department, Faculty of Sciences,
\\ University Ibn Zohr, P.O.Box 28/S\\ Agadir, Morocco.}
\date{}
\maketitle

\begin{abstract}
The so-called Gazeau-Klauder and Perelomov coherent states are introduced
for an arbitrary quantum system. We give also the general framework to
construct the generalized intelligent states which minimize the
Robertson-Schr\"odinger uncertainty relation. As illustration, the
P\"oschl-Teller potentials of trigonometric type will be chosen. We show the
advantage of the analytical representations of Gazeau-Klauder and Perelomov
coherent states in obtaining the generalized intelligent states in
analytical way.
\end{abstract}

\newpage\

\section{Introduction}

Coherent states, known as the closest states to classical ones,
play an important role in many different contexts of theoretical
and experimental physics, especially quantum optics $\left[
1,2,3\right] .$ Schr\"odinger first discovered the coherent states
for the harmonic oscillator potential in 1926 $\left[ 4\right] $
and much work has been done since then on their properties and
applications $\left[ 5,6\right] $. The coherent states have
also been found in systems with the Lie group symmetry $\left[ 7,8\right] $%
. Recently, coherent states have been found in special
Hamiltonians $\left[ 9\right] $. These coherent states are called
minimum uncertainty coherent states. In coherent states the
standard deviation of $X$ (coordinate) and $P$ (momentum) are
equal and their product is minimum over states. There are also
quantum states where, through we have minimum uncertainty for the
standard deviation of coordinate and momentum, they are not equal
any more; those states are called squeezed states. These states
are as important as coherent ones their generation play an
important role in many different branch of physics.

   There exist three definitions of coherent states. The first one defines the
usual coherent states as eigenstates of the annihilation operator $a^{-}$
for each individual oscillator mode of the electromagnetic field

\begin{equation}
a^{-}\left| z\right\rangle =z\left| z\right\rangle.
\end{equation}
Here $\left[ a^{-},a^{+}\right] =1$ ($\left( a^{-}\right) ^{\dagger }=a^{+}$%
) and $z$ is a complex constant with conjugate $\bar z$. The resulting unit
normalized states $\left| z\right\rangle $ are given by
\begin{equation}
\left| z\right\rangle =e^{-\frac{\left| z\right| ^2}2}\sum\limits_{n=0}^%
\infty \frac{z^n}{\sqrt{n!}}\left| n\right\rangle,
\end{equation}
where $\left| n\right\rangle $ is an element of the Fock space $\mathcal{H}%
\equiv \left\{ \left| n\right\rangle ,n\geq 0\right\} $. A second definition
of coherent states for oscillators assumes the existence of a unitary $%
^{\prime \prime }$displacement$^{\prime \prime }$ operator $D\left( z\right)
$ defined as

\begin{equation}
D\left( z\right) =\exp \left( za^{+}-\bar za^{-}\right).
\end{equation}
The coherent states parametrized by $z$ are given by the action of $D\left(
z\right) $ on the ground state $\left| 0\right\rangle $. The unitarity of $%
D\left( z\right) $ ensures the correct normalization of $\left|
z\right\rangle $. The Baker-Campbell-Hausdorff relation (BCH)
\begin{equation}
e^Ae^B=e^{A+B+\frac 12\left[ A,B\right] },
\end{equation}
valid only for any two operators $A$ and $B$ that both commute with the
commutator $\left[ A,B\right] $, implies the equivalence of this definition
with the one above.

A third definition is based on the uncertainty relation, with the position $X
$ and momentum $P$ given, as usual, by

\begin{equation}
X=\frac 1{\sqrt{2}}\left( a^{-}+a^{+}\right) ,\hspace{1cm}%
P=\frac i{\sqrt{2}}\left( a^{+}-a^{-}\right).
\end{equation}
The coherent states defined above have the minimum-uncertainty value $%
2\Delta X\Delta P=1$ and maintain this relation in time (temporal stability
of coherent states). Coherent states have two important properties. First,
they are not orthogonal to each other. Second, they provide a resolution of
the identity, i.e., they form an over complete set states.

A central goal of this article is to extend the above three
definitions for an arbitrary quantum system (exactly solvable) and
comparing the equivalence between them. Note that an attempt in
this sense was considered by Nieto et al $\left[ 9\right] $
concluding that the three definition are generally inequivalents.
Our analysis is different from the Nieto et al ones for several
reasons which will be clear in the sequel of this paper.

The method we adopt is an extension of the group-theoretical
approach to coherent states which generalizes the displacement
operator definition. We call the obtained coherent states:
coherent states of Perelomov type. The latter will be compared
with Gazeau-Klauder coherent states constructed using the approach
adopted by Barut-Girardello $\left[ 10,11\right] $ (see also the
references $12$ , $13$ and $14$) for an arbitrary quantum system.
To extend to third definition, we solve the eigenvalue equation of
states minimizing the Robertson-Schr\"odinger uncertainty relation
which extend the Heisenberg one. These states are called
Generalized Intelligent States (GIS) $\left[ 15,16\right] $. We
show that the set of GIS includes the Gazeau-Klauder coherent
states in a particular situation.

This paper is organized as follows: Creation and annihilation
operators for an arbitrary quantum system (exactly solvable) are
introduced in section 2. These operators are used to define
Gazeau-Klauder coherent states in section 3. Section 4 is devoted
to give a general algorithm leading to the Perelomov coherent
states. States minimizing the Robertson-Schr\"odinger uncertainty
relation are constructed in section 5. The results of sections 3,
4 and 5 are applied to a quantum system evolving in
P\"oschl-Teller potentials. In particular, using the analytical
representations of Gazeau-Klauder coherent states and Perelomov
ones, we give the generalized intelligent states under analytical
forms (section 6). The last section concerns a summary of the main
results of this work.

\section{Creation and annihilation operators for an arbitrary quantum system}

We start with general consideration on the creation and annihilation
operators from the factorization of a given Hamiltonian admitting a
non-degenerate discrete infinite energy spectrum. Let us assume that the
Hamiltonian $H$ of a quantum system admits infinite spectrum of energy $%
\left\{ E_n, n=0,1,2,...\right\} $ such that the fundamental energy $%
E_0=0$ and the others are in increasing order, i.e.,
\begin{equation}
E_0=0<E_1<E_2...<E_{n-1}<E_n<...
\end{equation}
For such a system, we known that the fundamental state $\psi
_0(x)$ and the potential $V(x)$ are closely related so that the
factorization is possible. Indeed, the time independent
Schr\"odinger equation for $\psi _0(x)$ reads
\begin{equation}
H\psi _0(x)=\left( -\frac 12\frac{d^2}{dx^2}+V(x)\right) \psi
_0(x)=0,
\end{equation}
and we have
\begin{equation}
V(x)=\frac 12\frac{\psi _0^{^{\prime \prime }}(x)}{\psi _0(x)},
\end{equation}
where the prime means the derivation with respect to $x.$

The usual factorization of $H$ is then given by
\begin{equation}
H=A^{+}A^{-}
\end{equation}
with
\begin{equation}
A^{+}=\frac 1{\sqrt{2}}\left( -\frac d{dx}+W(x)\right),%
\hspace{0.8cm}A^{-}=\frac 1{\sqrt{2}}\left( \frac
d{dx}+W(x)\right),
\end{equation}
where the superpotential $W(x)$ satisfies the Riccati equation
\begin{equation}
V(x)=\frac 12\left( W^2(x)-W^{^{\prime }}(x)\right).
\end{equation}
It is clear, from equations $\left( 8\right) $ and $\left( 11\right) $, that
$W(x)$ takes the form
\begin{equation}
W(x)=-\frac{\psi _0^{^{\prime }}(x)}{\psi _0(x)}.
\end{equation}
From equation $\left( 10\right) $, we have

\begin{equation}
\left[ A^{-},A^{+}\right] =W^{^{\prime }}(x),
\end{equation}
which generalizes the usual one for the harmonic oscillator ($W\left(
x\right) =x$). The operators $A^{+}$ and $A^{-}$ are not the creation and
annihilation operators of $H.$ Then, we are interested now in identifying
the operators creating and annihilating the quantum states of the system
under consideration. The key ingredients in constructing them is to define
the operator $H_{+}=A^{-}A^{+}$ obtained from $H=H_{-}=A^{+}A^{-}$ by
reversing the order of $A^{-}$ and $A^{+}.$ The operator $H_{+}$ is in fact
an Hamiltonian corresponding to a new potential $V_{+}\left( x\right) .$%
\begin{equation}
H_{+}=-\frac 12\frac{d^2}{dx^2}+V_{+}\left( x\right),%
\hspace{1cm}V_{+}(x)=\frac 12\left( W^2(x)+W^{^{\prime
}}(x)\right).
\end{equation}
The potentials $V_{-}\left( x\right) =V\left( x\right) $ and $V_{+}(x)$ are
known as supersymmetric partner potentials and $H_{-}\equiv H$ and $H_{+}$
are isospectrals ($H_{+}$ is also exactly solvable). Indeed, the
Schr\"odinger equation for $H_{-}$

\begin{equation}
H_{-}\left| \psi _n\right\rangle =E_n\left| \psi _n\right\rangle,
\end{equation}
implies
\begin{equation}
H_{+}\left( A^{-}\left| \psi _n\right\rangle \right) =E_n\left(
A^{-}\left| \psi _n\right\rangle \right).
\end{equation}
Similarly, the Schr\"odinger equation for $H_{+}$

\begin{equation}
H_{+}\left| \theta _n\right\rangle =e_n\left| \theta
_n\right\rangle,
\end{equation}
implies
\begin{equation}
H_{-}\left( A^{+}\left| \theta _n\right\rangle \right) =e_n\left(
A^{+}\left| \theta _n\right\rangle \right),
\end{equation}
where $e_n$ are the eigenvalues and $\left| \theta _n\right\rangle $ are
eigenstates of $H_{+}.$ From the latters equations and the fact that $E_0=0$%
, it is clear that the energies and eigenstates of $H_{-}$ and $H_{+}$ are
related by

\begin{eqnarray}
e_n &=&E_{n+1},  \nonumber \\ A^{-}\left| \psi_{n+1}\right\rangle
&=&\sqrt{E_{n+1}}e^{i\left(E_{n+1}-E_n\right)\alpha }\left|
\theta_n\right\rangle,\\ A^{+}\left|\theta_n\right\rangle
&=&\sqrt{e_n}e^{-i\left(E_{n+1}-E_n\right)\alpha}\left|\psi
_{n+1}\right\rangle,
\end{eqnarray}
where $\alpha \in \mathbf{R}$. Notice that if the eigenstates $\left| \psi
_{n+1}\right\rangle $ ($\left| \theta _n\right\rangle $) of $H_{-}$ ($H_{+}$%
) is normalized, then the wavefunctions $\left| \theta _n\right\rangle $ ($%
\left| \psi _{n+1}\right\rangle $) in equations $\left( 19\right) $ and $%
\left( 20\right) $ is also normalized. Further, the operator $A^{-}$ ($A^{+}$%
) converts an eigenfunction of $H_{-}$ ($H_{+}$) into an eigenfunction of $%
H_{+}$ ($H_{-}$) with the same energy. Thus, the operators $A^{-}$ and $A^{+}
$ connect the states $\left| \psi _n\right\rangle $ and $\left| \theta
_n\right\rangle $ and can not be considered as creation and annihilation
operators for $H\equiv H_{-}.$ To define the ladder operators for the
quantum system described by $H$, we consider the unitary transformation $U$
connecting the basis \{$\left| \psi _n\right\rangle $\} and \{$\left| \theta
_n\right\rangle $\} as follows
\begin{equation}
\left| \theta _n\right\rangle =U\left| \psi _n\right\rangle
\end{equation}
with
\begin{equation}
UU^{+}=U^{+}U=I.
\end{equation}
The explicit structure of the unitary operator $U$ is given by

\begin{equation}
U=\sum\limits_{n,m}U_{nm}\left| \psi _n\right\rangle \left\langle
\psi _m\right|,
\end{equation}
where the elements $U_{nm}$ are evaluated by

\begin{equation}
U_{nm}=\langle \psi _n\left| \theta _m\right\rangle =\int \psi
_n^{*}\left( x\right) \theta _m\left( x\right) dx.
\end{equation}
Note that in the harmonic oscillator case $U=I.$

At this stage, we can introduce the creation and annihilation operators of $%
H $ by

\begin{equation}
a^{+}=A^{+}U,\hspace{1cm}a^{-}=U^{+}A^{-}.
\end{equation}
The actions of the operators $a^{+}$ and $a^{-}$ on the states \{$\left|
\psi _n\right\rangle $\} are given by

\begin{eqnarray}
a^{+}\left| \psi _n\right\rangle &=&\sqrt{E_{n+1}}\hbox{
}e^{-i\left( E_{n+1}-E_n\right) \alpha }\left| \psi
_{n+1}\right\rangle, \\ a^{-}\left| \psi _n\right\rangle
&=&\sqrt{E_n}\hbox{ }e^{i\left( E_n-E_{n-1}\right) \alpha }\left|
\psi _{n-1}\right\rangle.
\end{eqnarray}
Note that $a^{+}a^{-}=A^{+}A^{-}=H.$ It is easy to show that
\begin{equation}
\left| \psi _n\right\rangle =\frac{\left( a^{+}\right) ^n}{\sqrt{E(n)}}%
e^{iE_n\alpha }\left| \psi _0\right\rangle,\hspace{1cm}n>0,
\end{equation}
where we have defined
\begin{equation}
E(n)=E_1E_2...E_n
\end{equation}
and for $n=0$ \hspace{0.2cm},\hspace{0.3cm} $E(0)=1.$

The exponential factor appearing in all these expressions produces only a
phase factor and will be significant for the temporal stability of the
coherent states we will construct in the following. From the equations $%
\left( 26\right) $ and $\left( 27\right) ,$ we have also
\begin{equation}
\left[ a^{-},a^{+}\right] \left| \psi _n\right\rangle =\left(
E_{n+1}-E_n\right) \left| \psi _n\right\rangle.
\end{equation}
Let us now introduce the operator $N$ such that

\begin{equation}
N\left| \psi _n\right\rangle =n\left| \psi _n\right\rangle,
\end{equation}
which in general (for an arbitrary quantum system) different from the
product $a^{+}a^{-}$ ($=H$). We can see that it satisfies the following
properties
\begin{equation}
a^{-}N=\left( N+1\right) a^{-},\hspace{0.8cm}a^{+}\left(
N+1\right) =Na^{+}.
\end{equation}
We are then able to define an operator $G$ such that
\begin{equation}
\left[ a^{-},a^{+}\right] =G(N),
\end{equation}
which acts in the states $\left| \psi _n\right\rangle $ as
\begin{equation}
G(N)\left| \psi _n\right\rangle =\left( E_{n+1}-E_n\right) \left|
\psi _n\right\rangle.
\end{equation}
The operator $G$ is hermitian.

\section{Gazeau-Klauder coherent states}

\subsection{Eigenstates of annihilation operator}

The Gazeau-Klauder coherent states are eigenstates of the annihilation
operator of the system under consideration. For the system governed by the
Hamiltonian $H$ ($=A^{+}A^{-}=a^{+}a^{-}$), such states are labelled by $%
\left| z,\alpha \right\rangle $, $z\in \mathbf{C}$ and $\alpha \in \mathbf{R}
$ ($\alpha $ is the parameter entering in the eqs $\left( 26\right) $ and $%
\left( 27\right) $), and they assumed to be the solution of the eigenvalue
equation
\begin{equation}
a^{-}\left| z,\alpha \right\rangle =z\left| z,\alpha
\right\rangle.
\end{equation}
To have their explicit form we decompose it in the basis $\left\{ \left|
\psi _n\right\rangle \right\} $ such that
\begin{equation}
\left| z,\alpha \right\rangle =\sum\limits_{n=0}^{+\infty
}a_n\left| \psi _n\right\rangle
\end{equation}
and insert this equation in $\left( 35\right) .$ Using equation $\left(
27\right) $, we find
\begin{equation}
a_n=\frac{z^n}{\sqrt{E(n)}}e^{-iE_n\alpha }a_0,\hspace{1cm}n>0,
\end{equation}
with $E(n)$ is given by $\left( 29\right) $. For $n=0$, the states $\left|
\psi _0\right\rangle $ is an eigenstate of $a^{-}$ with eigenvalue $0$.
Finally, the coherent states $\left| z,\alpha \right\rangle $ take the form
\begin{equation}
\left| z,\alpha \right\rangle =a_0\sum\limits_{n=0}^{+\infty }\frac{z^n}{%
\sqrt{E(n)}}e^{-iE_n\alpha }\left| \psi _n\right\rangle.
\end{equation}
The constant $a_0$ will be fixed by imposing the normalization to unity. We
get
\begin{equation}
\left| a_0\right| ^{-2}=\sum\limits_{n=0}^{+\infty }\frac{\left|
z\right| ^{2n}}{E(n)}.
\end{equation}
The coherent states $\left( 38\right) $ are continuous in $z\in \mathbf{C}$
and $\alpha \in \mathbf{R}$. Moreover, the presence of the phase factor in
the definitions equations $\left( 26\right) $ and $\left( 27\right) $ of the
$a^{-}$ and $a^{+}$ actions leads to temporal stability of the coherent
states. Indeed, we have
\begin{equation}
e^{iHt}\left| z,\alpha \right\rangle =\left| z,\alpha
+t\right\rangle.
\end{equation}
The analysis of completeness (in fact, the overcompleteness) require to
compute the identity resolution, that is
\begin{equation}
\int \left| z,\alpha \right\rangle \left\langle z,\alpha \right|
d\mu (z)=I_{\mathcal{H}}.
\end{equation}
Note that the integral is over the disk $\left\{ z\in \mathbf{C,}\hbox{ }%
\left| z\right| <\mathcal{R}\right\} $, where the radius of convergence $%
\mathcal{R}$ is
\begin{equation}
\mathcal{R}=\lim\limits_{n\rightarrow \infty }\sqrt[n]{E(n)}
\end{equation}
and the measure $d\mu (z)$ has to be determined. To determine it, we suppose
that $d\mu (z)$ depends only on $\left| z\right| $ (isotropy condition), we
take
\begin{equation}
d\mu (z)=\left[ a_0\right] ^{-2}h(r^2)rdrd\varphi \hspace{1cm};\hspace{1.5cm}%
z=re^{i\varphi }.
\end{equation}
Hence, the identity resolution can be written in the following form
\begin{equation}
I_{\mathcal{H}}=\sum\limits_{n=0}^{+\infty }\left| \psi
_n\right\rangle
\left\langle \psi _n\right| \left[ \frac \pi {E(n)}\int\nolimits_0^{%
\mathcal{R}^2}h(u)u^ndu\right].
\end{equation}
The last equation is satisfied when we have
\begin{equation}
\int\nolimits_0^{\mathcal{R}^2}h(u)u^ndu=\frac{E(n)}\pi.
\end{equation}
It is clear that the identity resolution is then equivalent to the
determination of the function $h(u)$ satisfying the equation $\left(
45\right) $. For $\mathcal{R}\rightarrow \infty $, the function $h(u)$ is
the inverse Mellin transform of $\pi ^{-1}E(s-1)$%
\begin{equation}
h(u)=\frac 1{2\pi ^2i}\int\nolimits_{c-i\infty }^{c+i\infty }E(s-1)u^{-s}ds%
;\hspace{1.5cm}c\in \mathbf{R}.
\end{equation}
Note that explicit computation of the function $h(u)$ requires the knowledge
of the spectrum of the quantum mechanical system under consideration.

Using the equation $\left( 35\right) $, one can obtain the mean value of the
Hamiltonian $H$ in the states $\left| z,\alpha \right\rangle $%
\begin{equation}
\left\langle z,\alpha \right| H\left| z,\alpha \right\rangle
=\left| z\right| ^2.
\end{equation}
This relation in known as the action identity.

Finally, we remark that the coherent states $\left| z,\alpha \right\rangle $
can be written as an operator $U(z)$ acting in the ground state $\left| \psi
_0\right\rangle $%
\begin{equation}
U(z)=a_0\exp \left( z\frac N{g(N)}a^{+}\right)
\end{equation}
such that we have
\begin{equation}
\left| z,\alpha \right\rangle =U(z)\left| \psi _0\right\rangle.
\end{equation}
In $\left( 48\right) $, $g(N)\equiv H=a^{+}a^{-}$. The operator $U(z)$ is
not unitary and cannot be interpreted as the displacement operator in the
Perelomov's sense.

A final comment can be made in connection with the work of Gazeau and
Klauder $\left[ 11\right] $. In fact, the coherent states $\left( 38\right) $
satisfy all the requirements (continuity, temporal stability, identity
resolution, action identity) given in their approach but they are more
general since we are working with $z\in \mathbf{C}$ and $\alpha \in \mathbf{R%
}$. They are eigenstates of the annihilation operator $a^{-}$. Additional
properties of this set of states will be considered in section $\left(
5\right) .$

\subsection{Fock-Bargmann representation}

It is well known that the Fock-Bargmann representation enable one to find
simple solutions of a number of problems, exploiting the theory of
analytical entire functions. In this subsection, generalizing the pioneering
work of Bargmann $\left[ 17\right] $ for the usual harmonic oscillator, we
give the Bargmann representation of an arbitrary quantum mechanical system.
We recall that in the Fock-Bargmann representation for the standard harmonic
oscillator, the creation operator $a^{+}$ is the multiplication by $z$ while
the annihilation operator $a^{-}$ is the differentiation with respect to $z$

For an arbitrary quantum system, we define the Fock-Bargmann space as a
space of functions which are holomorphic on a ring $D$ of the complex plane.
The scalar product is written with an integral of the form

\begin{equation}
\langle f\left| g\right\rangle =\int \overline{f(z)}g(z)d\mu (z),
\end{equation}
where $d\mu (z)$ is the measure defined above (see Eq $\left(
43\right) $). Let $\left| f\right\rangle $ be an arbitrary quantum
state of the system under study
\begin{equation}
\left| f\right\rangle =\sum\limits_{n=0}^{+\infty }f_n\left| \psi
_n\right\rangle \hspace{0.5cm},\hbox{with}\hspace{1cm}%
\sum\limits_{n=0}^{+\infty }\left| f_n\right| ^2<\infty.
\end{equation}
Any state $\left| f\right\rangle $ is represented, in the Fock-Bargmann
representation, as a function of the complex variable $z$ (using the
so-called coherent states associated with the quantum system under
consideration)
\begin{equation}
f(z)\equiv \langle \overline{z},\alpha \left| f\right\rangle
=\sum\limits_{n=0}^{+\infty }\frac{z^n}{\sqrt{E(n)}}e^{iE_n\alpha
}f_n.
\end{equation}
In particular, to the vectors $\left| \psi _n\right\rangle $ there
correspond the monomials

\begin{equation}
\langle \overline{z},\alpha \left| \psi _n\right\rangle =\frac{z^n}{\sqrt{%
E(n)}}e^{iE_n\alpha }.
\end{equation}
Using the equations $\left( 52\right) $ and $\left( 53\right) $, we can
prove the following result: In the Fock-Bargmann representation, we realize
the annihilation operator $a^{-}$ by
\begin{equation}
a^{-}=z^{-1}g(z\frac d{dz}),
\end{equation}
the creation operator $a^{+}$%
\begin{equation}
a^{+}=z,
\end{equation}
and the operator number by
\begin{equation}
N=z\frac d{dz}.
\end{equation}
The Fock-Bargmann representation exists if we have a measure such that

\begin{equation}
\int \left| z,\alpha \right\rangle \left\langle z,\alpha \right|
d\mu (z)=I_{\mathcal{H}}.
\end{equation}
The existence of the measure, discussed previously for the so-called
Gazeau-Klauder coherent states, ensures that the scalar product takes the
form $\left( 50\right) $. We note that in the case where
\begin{equation}
g(z\frac d{dz})=z\frac d{dz},\hbox{i.e.}\hspace{1cm}g(N)=N.
\end{equation}
we recover the well-known Fock-Bargmann representation of the harmonic
oscillator. The Fock-Bargmann realization discussed here will be the main
tool to construct the generalized intelligent states (see section 6).

\section{Coherent states of Perelomov's type}

In view of the second definition of coherent states for the standard
harmonic oscillator (group-theoretical approach), we define, for an
arbitrary quantum system, the states
\begin{equation}
\left| z,\alpha \right\rangle =\exp \left( za^{+}-\overline{z}a^{-}\right)
\left| \psi _0\right\rangle ,\hbox{for}\hspace{1cm}z\in \mathbf{C,%
}
\end{equation}
which we call of Perelomov's type. We have to compute the action of the
displacement operator
\begin{equation}
D(z)=\exp \left( za^{+}-\overline{z}a^{-}\right)
\end{equation}
on the ground state $\left| \psi _0\right\rangle $ of the quantum
system under study. We will give the result of this action in a
closed form. An illustration is treated for the P\"oschl-Teller
and square-well potentials (in section 6).

Using the actions of the annihilation and creation operators on the Hilbert
space $\left\{ \left| \psi _n\right\rangle ,n=0,1,2,...\right\} $ (eqs $%
\left( 26\right) $ and $\left( 27\right) $), one can, after more or less
complicated computations, show that the states $\left| z,\alpha
\right\rangle $ can be written as follows
\begin{equation}
\left| z,\alpha \right\rangle =\sum\limits_{n=0}^{+\infty }\frac{z^n}{\sqrt{%
F_n(\left| z\right| )}}e^{-iE_n\alpha }\left| \psi
_n\right\rangle.
\end{equation}
The quantities $F_n(\left| z\right| )$ satisfy
\begin{equation}
F_n(\left| z\right| )E(n)\left( c_n\left( \left| z\right| \right)
\right) ^2=1,
\end{equation}
where the coefficients $c_n\left( \left| z\right| \right) $ are given by
\begin{equation}
c_n\left( \left| z\right| \right) =\sum\limits_{j=0}^{+\infty }\frac{%
(-\left| z\right| ^2)^j}{\left( n+2j\right) !}\left(
\sum\limits_{i_1=1}^{n+1}E_{i_1}\sum\limits_{i_2=1}^{i_1+1}E_{i_2}....%
\sum\limits_{i_j=1}^{i_{j-1}+1}E_{i_j}\right).
\end{equation}
Setting
\begin{equation}
\pi \left( n+1,j\right)
=\sum\limits_{i_1=1}^{n+1}E_{i_1}\sum\limits_{i_2=1}^{i_1+1}E_{i_2}....%
\sum\limits_{i_j=1}^{i_{j-1}+1}E_{i_j}\hspace{0.3cm}\hbox{and}%
\hspace{0.2cm}\pi \left( n+1,0\right) =1,
\end{equation}
one can verify that the $\pi $'s satisfy the following relation:
\begin{equation}
\frac{\pi \left( n+1,j\right) -\pi \left( n,j\right)
}{E_{n+1}}=\pi \left( n+2,j-1\right).
\end{equation}
Using this recurrence formula, one can show that the coefficients $c_n\left(
\left| z\right| =r\right) $ satisfy the following differential equation
\begin{equation}
\frac{dc_n\left( r\right) }{dr}=\frac 1rc_{n-1}\left( r\right)
-\frac nrc_n\left( r\right) -E_{n+1}c_{n+1}\left( r\right) r.
\end{equation}
Hence, solving this equation, we can obtain explicitly the coherent states $%
\left| z,\alpha \right\rangle $ of Perelomov's type. Of course, to
solve this equation for an arbitrary quantum system is, in
general, not an easy task. However, solutions in some particular
(and interesting physical system) will be given in section $6$.
Here, as a first illustration of the approach leading to coherent
states of Perelomov's type, we give the standard harmonic
oscillator coherent states using the above considerations. In this
case we show that $\left( 61\right) $ coincides with $\left(
2\right) $. For the harmonic oscillator $E_n=n$ and $E\left(
n\right) =n!$.

To solve the equation $\left( 66\right) $, we set
\begin{equation}
c_n\left( r\right) =\frac 1{n!}\sum\limits_{m=0}^{+\infty }a_mr^m.
\end{equation}
Substituting this expression in $\left( 66\right) $, we get the coefficients
$a_m$,%
\begin{equation}
a_{2p}=\frac{\left( -1\right) ^p}{2^pp!}a_0\hspace{1cm}\hbox{and}%
\hspace{1cm}a_{2p+1}=0,
\end{equation}
where $a_0=1$ because $c_0\left( r=0\right) =1$. Finally, we have
\begin{equation}
F_n\left( \left| z\right| \right) =n!\exp \left( \left| z\right| ^2\right)
\end{equation}
and
\begin{equation}
\left| z,\alpha \right\rangle =\exp \left( -\frac{\left| z\right| ^2}%
2\right) \sum\limits_{n=0}^{+\infty
}\frac{z^n}{\sqrt{n!}}e^{-i\alpha n}\left| n\right\rangle.
\end{equation}
We recover a well known result.

\section{Generalized intelligent states}

These states minimize the Robertson-Schr\"odinger uncertainty relation $%
\left[ 18, 19\right] $, and generalize the Gazeau-Klauder coherent
states.

Using the creation $a^{+}$ and annihilation $a^{-}$ operators, we introduce
the hermitian operators
\begin{equation}
X=\frac 1{\sqrt{2}}\left( a^{+}+a^{-}\right), \hspace{1.6cm}P=\frac i{\sqrt{2}%
}\left( a^{+}-a^{-}\right),
\end{equation}
which satisfy the commutation relation
\begin{equation}
\left[ X,P\right] =iG\left( N\right) \equiv iG.
\end{equation}
The operator $G\left( N\right) $, defined by $\left( 34\right) $, is not
necessarily a multiple of the unit operator (for an arbitrary quantum
system). It is well known that for two hermitian operators $X$ and $P$
satisfying the noncanonical commutation relation $\left( 72\right) $, the
variances $(\Delta X)^2$ and $(\Delta P)^2$ satisfy the
Robertson-Schr\"odinger uncertainty relation
\begin{equation}
\left( \Delta X\right) ^2\left( \Delta P\right) ^2\geq \frac
14\left( \left\langle G\right\rangle ^2+\left\langle
F\right\rangle ^2\right),
\end{equation}
where the operator $F$ is defined by
\begin{equation}
F=\left\{ X-\left\langle X\right\rangle ,P-\left\langle P\right\rangle
\right\}
\end{equation}
or by
\begin{equation}
F=i\left[ \left( 2a^{-}-\left\langle a^{-}\right\rangle \right) \left\langle
a^{-}\right\rangle +\left( -2a^{+}+\left\langle a^{+}\right\rangle \right)
\left\langle a^{+}\right\rangle -a^{-2}+a^{+2}\right]
\end{equation}
in terms of the operators $a^{-}$ and $a^{+}$. The symbol $\left\{ ,\right\}
$ in $\left( 74\right) $ stands for the anti-commutator. When there is a
correlation between $X$ and $P,$ i.e. $\left\langle F\right\rangle \neq 0$,
the relation $\left( 73\right) $ is a generalization of the usual one ( the
Heisenberg uncertainty condition)
\begin{equation}
\left( \Delta X\right) ^2\left( \Delta P\right) ^2\geq \frac
14\left\langle G\right\rangle ^2.
\end{equation}
The special form $\left( 76\right) $ is identical with the general form $%
\left( 73\right) $ if $X$ and $P$ are uncorrelated, i.e., $\left\langle
F\right\rangle =0$. The general uncertainty relation $\left( 73\right) $ is
better suited to determine the lower bound on the product of variances in
the measurement of observables corresponding to the noncanonical operators.
The so-called generalized intelligent states are obtained when the equality
in the Robertson-Schr\"odinger uncertainty relation is realized $\left[
20\right] $. The inequality in $\left( 73\right) $ becomes equality for the
states satisfying the equation (see also, $\left[ 20-23\right] $)
\begin{equation}
\left( X+i\lambda P\right) \left| \psi \right\rangle
=z\sqrt{2}\left| \psi \right\rangle, \hspace{1.5cm}\lambda ,z\in
\mathbf{C}.
\end{equation}
As a consequence, we have the following relations
\begin{equation}
\left( \Delta X\right) ^2=\left| \lambda \right| \Delta ,%
\hspace{1cm}\left( \Delta P\right) ^2=\frac 1{\left| \lambda
\right| }\Delta,
\end{equation}
with
\begin{equation}
\Delta =\frac 12\sqrt{\left\langle G\right\rangle ^2+\left\langle
F\right\rangle ^2}.
\end{equation}
The average values $\left\langle G\right\rangle $ and
$\left\langle F\right\rangle $, in the states satisfying the
eigenvalue equation $\left( 77\right) $, can be expressed in terms
of the variances as follows:
\begin{equation}
\left\langle G\right\rangle =2\hbox{Re}\left( \lambda \right)
\left\langle
\Delta P\right\rangle ^2,\hspace{1.5cm}\left\langle F\right\rangle =2\hbox{Im}%
\left( \lambda \right) \left\langle \Delta P\right\rangle ^2.
\end{equation}
It is clear, from $\left( 78\right) $, that if $\left| \lambda \right| =1$
we have
\begin{equation}
\left( \Delta X\right) ^2=\left( \Delta P\right) ^2.
\end{equation}
We call the states satisfying $\left( 81\right) $ with $\left| \lambda
\right| =1$, the generalized coherent states. For $\left| \lambda \right|
\neq 1$, the states are called generalized squeezed states.

Using Eq. $\left( 77\right) $, one can obtain some general
relations for the average values and dispersions of $X$ and $P$ in
the
states which minimize the Robertson-Schr\"odinger uncertainty relation $%
\left( 73\right) $. We have
\begin{eqnarray}
\left( \Delta X\right) ^2 &=&\frac 12\left( \hbox{Re}\left(
\lambda \right) \left\langle G\right\rangle +\hbox{Im}\left(
\lambda \right) \left\langle F\right\rangle \right),  \\
\left( \Delta P\right) ^2 &=&\frac 1{2\left| \lambda \right| ^2}\left( \hbox{%
Re}\left( \lambda \right) \left\langle G\right\rangle
+\hbox{Im}\left( \lambda \right) \left\langle F\right\rangle
\right), \\
\hbox{Im}\left( \lambda \right) \left\langle G\right\rangle  &=&\hbox{Re}%
\left( \lambda \right) \left\langle F\right\rangle.
\end{eqnarray}
In order to give a complete classification of the so-called
generalized intelligent states for an arbitrary quantum system, we
have to solve the eigenvalue equation $\left( 77\right) $. Such
computation was considered previously by the authors in $\left[
15, 16\right] $. The states minimizing the Robertson-Schr\"odinger
uncertainty relation are given by
\begin{equation}
\left| \psi \right\rangle \equiv \left| z,\lambda ,\alpha \right\rangle
=\sum\limits_{n=0}^{+\infty }d_n\left| \psi _n\right\rangle, \hspace{1.5cm}%
d_n\equiv d_n\left( z,\alpha ,\lambda \right).
\end{equation}
For the case where $\lambda \neq -1,$ the coefficients $d_n$ are given by
the following expression
\begin{equation}
d_n=d_0\frac{\left( 2z\right) ^n}{\left( 1+\lambda \right)
^n\sqrt{E\left( n\right) }}\left[ \sum\limits_{h=0\left( 1\right)
\left[ \frac n2\right] }\left( -1\right) ^h\frac{\left( 1-\lambda
^2\right) ^h}{\left( 2z\right) ^{2h}}\Delta \left( n,h\right)
\right] e^{-i\alpha E_n},
\end{equation}
where the symbol $\left[ \frac n2\right] $ stands for the integer part of $%
\frac n2$ and the function $\Delta \left( n,h\right) $ is defined
by

\begin{equation}
\Delta \left( n,h\right) =\sum\limits_{j_1=1}^{n-\left(
2h-1\right) }E_{j_1}\left[ \sum\limits_{j_2=j_1+2}^{n-\left(
2h-3\right) }E_{j_2}...\left[ ...\left[
\sum\limits_{j_h=j_{h-1}+2}^{n-1}E_{j_h}\right] \right]
...\right].
\end{equation}
We note that the case $\lambda =-1,$ leading to the unnormalized solution,
is not of interest.

The states $\left| z,\lambda ,\alpha \right\rangle $ can be also
given as the action of some operator on the ground state $\left|
\psi _0\right\rangle $ of $H$. A more or less complicated
manipulation give the following result:
\begin{equation}
\left| z,\lambda ,\alpha \right\rangle =U\left( \lambda ,z\right)
\left| \psi _0\right\rangle,
\end{equation}
where
\begin{equation}
U\left( \lambda ,z\right) =d_0\sum\limits_{n=0}^\infty \left( \left( \frac{%
2z}{\lambda +1}\right) \frac{a^{+}}{g\left( N\right) }+\left(
\frac{\lambda -1}{\lambda +1}\right) \frac 1{g\left( N\right)
}\left( a^{+}\right) ^2\right) ^n.
\end{equation}
Note that the states $\left| z,\lambda ,\alpha \right\rangle $ are stable
temporally. As a first illustration of this construction, we can obtain the
generalized intelligent states for the standard harmonic oscillator ($%
g\left( N\right) =N$). We have (up to normalization constant)
\begin{equation}
\left| z,\lambda ,\alpha \right\rangle =\exp \left[ \left( \frac{\lambda -1}{%
\lambda +1}\right) \frac{\left( a^{+}\right) ^2}2\right] \exp
\left[ \left( \frac{2z}{\lambda +1}\right) a^{+}\right] \left|
0\right\rangle,
\end{equation}
where $\left| 0\right\rangle $ is the ground states for the harmonic
oscillator.

The Gazeau-Klauder coherent states correspond to the situation $\lambda =1.$
In this case, the coefficients $d_n$ are given by
\begin{equation}
d_n=d_0\frac{z^n}{\sqrt{E\left( n\right) }}e^{-i\alpha E_n},
\end{equation}
and the coherent states $\left| z,\lambda =1,\alpha \right\rangle $ coincide
with Gazeau-Klauder ones $\left| z,\alpha \right\rangle $ given by eq $%
\left( 38\right) $. The normalization factor $d_0$ is given by eq $\left(
39\right) $. The states $\left| z,\lambda =1,\alpha \right\rangle \equiv
\left| z,\alpha \right\rangle $ minimize the Heisenberg uncertainty relation
$\left( 76\right) $ and are eigenvectors of the annihilation operator $a^{-}$%
. We have
\begin{equation}
\left( \Delta X\right) ^2=\left( \Delta P\right) ^2=\frac
12\left\langle G\right\rangle,
\end{equation}
where
\begin{equation}
\left\langle G\right\rangle =d_0^2\sum\limits_{n=0}^{+\infty
}\frac{\left|
z\right| ^{2n}}{E\left( n\right) }E_{n+1}-\left| z\right| ^2\hspace{1cm}%
\hbox{and}\hspace{1cm}\left\langle F\right\rangle =0.
\end{equation}
The latter equation traduce the fact that there is no correlation between $X$
and $P$. For the harmonic oscillator, it is easy to see that $\left\langle
G\right\rangle =1$ and $2\left( \Delta X\right) ^2=2\left( \Delta P\right)
^2=1$.

As we mentioned above, the coherent states minimizing
Robertson-Schr\"odinger uncertainty relation correspond to the case $\left|
\lambda \right| =1$. The case $\lambda =1$ correspond the Gazeau-Klauder
coherent states and $\lambda =-1$ is not allowed by our construction.
Setting $\lambda =e^{i\theta }$ ($\theta \neq k\pi ;k\in \mathbf{N}$), the
states $\left| z,\lambda ,\alpha \right\rangle $ are coherent and
dispersions are given by

\begin{equation}
\left( \Delta X\right) ^2=\left( \Delta P\right) ^2=\frac
1{2\left| \cos \theta \right| }\left\langle G\right\rangle.
\end{equation}
The main value of the operator $F$ is nonvanishing (vanish only in the
Gazeau-Klauder coherent states, i.e., $\lambda =1$) and it is given by
\begin{equation}
\left\langle F\right\rangle =\hbox{tg}\theta \left\langle
G\right\rangle.
\end{equation}
From the latter equation, we conclude that the presence of the correlation ($%
\left\langle F\right\rangle \neq 0$) does not forbid the system to
be prepared in a coherent states. This result is true for any
quantum system. The properties of the states $\left| z,\lambda
,\alpha \right\rangle $ turned out to be sensitive about the
spectral properties of the commutator $\left[ a^{-},a^{+}\right]
=G\left( N\right) $.

To close this section, we note that the minimization of the
Robertson-Schr\"odinger uncertainty relation leads to more general
expressions of coherent states associated to an arbitrary quantum system.
The Gazeau-Klauder coherent states ($\lambda =1$) (eigenvectors of the
annihilation operator) constitute a particular case of such coherent states
class ($\left| \lambda \right| =1$).

\section{Application: P\"oschl-Teller potentials}

We start by recalling the eigenvalues and eigenstates of infinite square
well and P\"oschl-Teller potentials $\left[ 24\right] $ (see also $\left[
25\right] $ and references therein). We consider the Hamiltonian
\begin{equation}
H=-\frac{d^2}{dx^2}+V_{\kappa ,\kappa ^{^{\prime }}}\left( x\right)
\end{equation}
describing a particle on the line, and submitted to the potential
\begin{equation}
V_{\kappa ,\kappa ^{^{\prime }}}(x)=\left\{
\begin{array}{c}
\frac 1{4a^2}\left[ \frac{\kappa \left( \kappa -1\right) }{\sin ^2\left(
\frac x{2a}\right) }+\frac{\kappa ^{^{\prime }}(\kappa ^{^{\prime }}-1)}{%
\cos ^2\left( \frac x{2a}\right) }\right] -\frac{(\kappa +\kappa
^{^{\prime }})^2}{4a^2},\hspace{0.6cm}0<x<\pi a \\ \infty
\hspace{5cm}x\leq 0,\hspace{0.5cm}x\geq \pi a
\end{array}
\right.
\end{equation}
for $\kappa >1$ and $\kappa ^{^{\prime }}>1$. It is well known that the
P\"oschl-Teller potentials interpolate between the harmonic oscillator and
infinite square well. The infinite square well takes place in the limit $%
\kappa =\kappa ^{^{\prime }}=1.$

The Hamiltonian $H$ can be written in the factorized form

\begin{equation}
H=A_{\kappa ,\kappa ^{^{\prime }}}^{+}A_{\kappa ,\kappa ^{^{\prime
}}}^{-},
\end{equation}
where the operators $A_{\kappa ,\kappa ^{^{\prime }}}^{-}$ and $A_{\kappa
,\kappa ^{^{\prime }}}^{+}$ are given by

\begin{equation}
A_{\kappa ,\kappa ^{^{\prime }}}^{\pm }=\mp \frac d{dx}+W_{\kappa ,\kappa
^{^{\prime }}}\left( x\right)
\end{equation}
in terms of the superpotentials $W_{\kappa ,\kappa ^{^{\prime }}}\left(
x\right) $%
\begin{equation}
W_{\kappa ,\kappa ^{^{\prime }}}\left( x\right) =\frac 1{2a}\left[
\kappa \hbox{cotg}\left( \frac x{2a}\right) -\kappa ^{^{\prime
}}\hbox{tang}\left( \frac x{2a}\right) \right].
\end{equation}
The eigenvectors are given by

\begin{equation}
\psi _n\left( x\right) =\left[ c_n(\kappa ,\kappa ^{^{\prime }})\right]
^{-\frac 12}\left( \cos \frac x{2a}\right) ^{\kappa ^{^{\prime }}}\left(
\sin \frac x{2a}\right) ^\kappa P_n^{(\kappa -\frac 12,\kappa ^{^{\prime
}}-\frac 12)}\left( \cos \left( \frac xa\right) \right)
\end{equation}
with $c_n(\kappa ,\kappa ^{^{\prime }})$ are the normalization constant
which takes the form
\begin{equation}
c_n\left( \kappa ,\kappa ^{^{\prime }}\right) =a\frac{\Gamma (n+\kappa
+\frac 12)\Gamma (n+\kappa ^{^{\prime }}+\frac 12)}{\Gamma (n+1)\Gamma
(n+\kappa +\kappa ^{^{\prime }})\Gamma (2n+\kappa +\kappa ^{^{\prime }})}
\end{equation}
and $P_n^{(\alpha ,\beta )}$'s stands for the Jacobi polynomials.

The eigenvalues of $H$ are given by
\begin{equation}
H\left| \psi _n\right\rangle =n(n+\kappa +\kappa ^{^{\prime
}})\left| \psi _n\right\rangle.
\end{equation}
To find the annihilation and creation operators for the P\"oschl-Teller
system, we follow the strategy given in section 2. So, we denote $H$ by $%
H_{-}$ and $V_{\kappa ,\kappa ^{\prime }}\left( x\right) $ by $V_{\kappa
,\kappa ^{\prime }}^{-}\left( x\right) $ the Hamiltonian $H_{+}=A_{\kappa
,\kappa ^{\prime }}^{-}A_{\kappa ,\kappa ^{\prime }}^{+}$ (supersymmetric
partner of $H\equiv H_{-}$)
\begin{equation}
H_{+}=-\frac 12\frac{d^2}{dx^2}+V_{\kappa ,\kappa ^{\prime
}}^{+}\left( x\right),
\end{equation}
describes a quantum system trapped in the potentials

\begin{equation}
V_{\kappa ,\kappa ^{^{\prime }}}(x)=\left\{
\begin{array}{c}
\frac 1{8a^2}\left[ \frac{\kappa \left( \kappa -1\right) }{\sin ^2\left(
\frac x{2a}\right) }+\frac{\kappa ^{^{\prime }}(\kappa ^{^{\prime }}-1)}{%
\cos ^2\left( \frac x{2a}\right) }\right] -\frac{(\kappa +\kappa
^{^{\prime }})^2}{8a^2},\hspace{0.6cm}0<x<\pi a \\
0,\hspace{5cm}x\leq 0,\hspace{0.5cm}x\geq \pi a.
\end{array}\right.
\end{equation}
The eigenstates of $H_{+}$ are given by

\begin{equation}
\theta _n\left( x\right) =\left[ c_n(\kappa +1,\kappa ^{^{\prime
}}+1)\right] ^{-\frac 12}\left( \cos \frac x{2a}\right) ^{\kappa
^{^{\prime }}+1}\left( \sin \frac x{2a}\right) ^{\kappa
+1}P_n^{(\kappa +\frac 12,\kappa ^{^{\prime }}+\frac 12)}\left(
\cos \left( \frac xa\right) \right),
\end{equation}
where the $c_n(\kappa ,\kappa ^{^{\prime }})$ are defined by $\left(
102\right) .$

The eigenvalues are $e_n=(n+1)(n+\kappa +\kappa ^{\prime }+1).$
Using the operators $A_{\kappa ,\kappa ^{\prime }}^{-}$ and
$A_{\kappa ,\kappa ^{\prime }}^{+}$ and the unitary transformation
$U$ connecting $\psi _n\left( x\right) $ and $\theta _n\left(
x\right) $ (see section 2), we define the creation and
annihilation operators by

\begin{equation}
a_{\kappa ,\kappa ^{\prime }}^{+}=A_{\kappa ,\kappa ^{\prime }}^{+}U%
\hspace{0.6cm}\hbox{and }\hspace{0.6cm}a_{\kappa ,\kappa ^{\prime
}}^{-}=U^{+}A_{\kappa ,\kappa ^{\prime }}^{-}.
\end{equation}
The creation and annihilation operators $a_{\kappa ,\kappa ^{^{\prime
}}}^{+} $ and $a_{\kappa ,\kappa ^{^{\prime }}}^{-}$ act on $\left| \psi
_n\right\rangle $ as follows

\begin{eqnarray}
a_{\kappa ,\kappa ^{^{\prime }}}^{+}\left| \psi _n\right\rangle &=&\sqrt{%
\left( n+1\right) \left( n+1+\kappa +\kappa ^{^{\prime }}\right) }%
e^{-i\alpha (2n+1+\kappa +\kappa ^{^{\prime }})}\left| \psi
_{n+1}\right\rangle  \nonumber  \label{nb}, \\
a_{\kappa ,\kappa ^{^{\prime }}}^{-}\left| \psi _n\right\rangle &=&\sqrt{%
n\left( n+\kappa +\kappa ^{^{\prime }}\right) }e^{i\alpha
(2n-1+\kappa +\kappa ^{^{\prime }})}\left| \psi
_{n-1}\right\rangle,
\end{eqnarray}
and satisfy the following commutation relation
\begin{equation}
\left[ a_{\kappa ,\kappa ^{^{\prime }}}^{-},a_{\kappa ,\kappa
^{^{\prime }}}^{+}\right] =G_{\kappa ,\kappa ^{^{\prime }}}\left(
N\right),
\end{equation}
where
\begin{equation}
G_{\kappa ,\kappa ^{^{\prime }}}\left( N\right) \equiv G\left(
N\right) =2N+(1+\kappa +\kappa ^{^{\prime }}).
\end{equation}
We note that $N\neq a_{\kappa ,\kappa ^{^{\prime }}}^{+}a_{\kappa ,\kappa
^{^{\prime }}}^{-}=H.$

\subsection{Gazeau-Klauder coherent states}

Using the result of section $\left( 3\right) $, the so-called Gazeau-Klauder
coherent states (eigenstates of the annihilation operator $a_{\kappa ,\kappa
^{^{\prime }}}^{-}$) reads as

\begin{equation}
\left| z,\alpha \right\rangle =\mathcal{N}\left( \left| z\right|
\right) \sum\limits_{n=0}^{+\infty }\frac{z^ne^{-i\alpha
n(n+\kappa +\kappa
^{^{\prime }})}}{\sqrt{\Gamma (n+1)\Gamma (n+\kappa +\kappa ^{^{\prime }}+1)}%
}\left| \psi _n\right\rangle  \label{tuyg},
\end{equation}
with $\mathcal{N}\left( \left| z\right| \right) $ the normalization constant
which takes the form
\begin{equation}
\left[ \mathcal{N}\left( \left| z\right| \right) \right]
^2=\frac{\left| z\right| ^{\kappa +\kappa ^{^{\prime
}}}}{I_{\kappa +\kappa ^{^{\prime }}}\left( 2\left| z\right|
\right) }  \label{RFD},
\end{equation}
where $I_{\kappa +\kappa ^{^{\prime }}}\left( 2\left| z\right| \right) $ is
the modified Bessel function of the first kind.

The identity resolution is given explicitly by
\begin{equation}
\int \left| z,\alpha \right\rangle \left\langle z,\alpha \right|
d\mu \left( z\right) =I_{\mathcal{H}},
\end{equation}
where the measure can be computed by the inverse Mellin transform $\left[
26\right] $

\begin{equation}
d\mu \left( z\right) =\frac 2\pi I_{\kappa +\kappa ^{^{\prime }}}\left(
2r\right) K_{\frac{\kappa +\kappa ^{^{\prime }}}2}\left( 2r\right) rdrd\phi, %
\hspace{1.5cm}z=r^{i\phi}.
\end{equation}
The Gazeau-Klauder coherent states of the infinite square well are obtained
from the P\"oschl-Teller ones simply by putting $\kappa +\kappa ^{^{\prime
}}=2.$

The Gazeau-Klauder coherent states form an overcomplete family of states
(resolving the unity by integration with respect to the measure given by $%
\left( 114\right) $), and provide a representation of any state $\left|
f\right\rangle $ by an entire function
\begin{eqnarray}
f(z,\alpha ) &=&\sqrt{\frac{I_{\kappa +\kappa ^{^{\prime }}}\left(
2\left| z\right| \right) }{\left| z\right| ^{\kappa +\kappa
^{^{\prime }}}}}\langle \overline{z},\alpha \left| f\right\rangle
\nonumber \\ \  &=&\sum\limits_{n=0}^{+\infty }\langle \psi
_n\left| f\right\rangle \frac{z^ne^{i\alpha n(n+\kappa +\kappa
^{^{\prime }})}}{\sqrt{\Gamma \left( n+1\right) \Gamma \left(
n+\kappa +\kappa ^{^{\prime }}+1\right) }}.
\end{eqnarray}
In particular, the analytic functions corresponding to the vectors $\left|
\psi _n\right\rangle $ are
\begin{equation}
\mathcal{F}_n\left( z,\alpha \right) =\frac{z^ne^{i\alpha
n(n+\kappa +\kappa ^{^{\prime }})}}{\sqrt{\Gamma \left( n+1\right)
\Gamma \left( n+\kappa +\kappa ^{^{\prime }}+1\right) }}.
\end{equation}
Using the Fock-Bargmann representation discussed in the subsection $\left(
3.2\right) $, the creation and annihilation operators, for quantum system
evolving in P\"oschl-Teller (or in the infinite square well) potentials, are
realized by
\begin{equation}
a_{\kappa ,\kappa ^{^{\prime }}}^{+}=z,\hspace{0.6cm}a_{\kappa
,\kappa ^{^{\prime }}}^{-}=z\frac{d^2}{dz^2}+(\kappa +\kappa
^{^{\prime }}+1)\frac d{dz},
\end{equation}
and the operator $G_{\kappa ,\kappa ^{^{\prime }}}\left( N\right) $, in this
representation, acts as
\begin{equation}
G=2z\frac d{dz}+(\kappa +\kappa ^{^{\prime }}+1).
\end{equation}
In fact, one can verify that
\begin{eqnarray}
a_{\kappa ,\kappa ^{^{\prime }}}^{+}\mathcal{F}_n\left( z,\alpha \right)  &=&%
\sqrt{\left( n+1\right) \left( n+1+\kappa +\kappa ^{^{\prime }}\right) }%
e^{-i\alpha (2n+1+\kappa +\kappa ^{^{\prime
}})}\mathcal{F}_{n+1}\left( z,\alpha \right)\\
a_{\kappa ,\kappa ^{^{\prime }}}^{-}\mathcal{F}_n\left( z,\alpha \right)  &=&%
\sqrt{n\left( n+\kappa +\kappa ^{^{\prime }}\right) }e^{i\alpha
(2n-1+\kappa +\kappa ^{^{\prime }})}\mathcal{F}_{n-1}\left(
z,\alpha \right),\\ G_{\kappa ,\kappa ^{^{\prime }}}\left(
N\right) \mathcal{F}_n\left( z,\alpha \right)  &=&(2n+1+\kappa
+\kappa ^{^{\prime }})\mathcal{F}_n\left( z,\alpha \right).
\end{eqnarray}
This realization will be useful, as we will see, to construct the
P\"oschl-Teller generalized intelligent states which minimize the
Robertson-Schr\"odinger uncertainty relation.

\subsection{P\"oschl-Teller coherent states of Perelomov's type}

In section $\left( 4\right) $, we defined coherent states of Perelomov's
type for an arbitrary quantum system. The expressions of these states are
given by infinite series (more or less complicated). As a first
illustration, we discussed the harmonic oscillator system. Here, we
construct the P\"oschl-Teller coherent states \`a la Perelomov. In this
order, we have to solve the differential equation $\left( 66\right) $ for
the P\"oschl-Teller potentials ($E_n=n(n+\kappa +\kappa ^{^{\prime }})$) .
In this case, the solutions are

\begin{equation}
c_n(r)=\frac 1{n!r^n}\ss _{m,n+\frac 12(\kappa +\kappa ^{^{\prime
}}+1)}^{-\frac 12(\kappa +\kappa ^{^{\prime }}+1)}\left( \cosh
(2r)\right),
\end{equation}
because the Jacobi functions $\ss $ satisfy the following differential
equation $\left[ 27\right] $
\begin{equation}
\frac d{dr}\ss _{m,n-l}^l(\cosh \left( 2r\right) )=n\ss
_{m,n-1-l}^l(\cosh \left( 2r\right) )-\left( n-2l\right) \ss
_{m,n+1-l}^l(\cosh \left( 2r\right) ),
\end{equation}
where $l=-\frac 12(\kappa +\kappa ^{^{\prime }}+1)$ and $m$ is a free
integer parameter which will be fixed after. These functions play an
important role in the representation theory of the $QU(2)$ group of
unimodular quasi-unitary matrices.

The differential equation $\left( 123\right) $ admits several solutions.
However, an admissible solution is obtained by noting that $D\left(
z=0\right) =\mathbf{1}$. Using the definition of the Jacobi functions $%
\left[ 27\right] ,$The unique solution, compatible with the condition $%
D\left( z=0\right) =\mathbf{1,}$ is given by
\begin{equation}
c_n(r)=\frac 1{n!r^n}\ss _{\frac 12(\kappa +\kappa ^{^{\prime
}}+1),n+\frac 12(\kappa +\kappa ^{^{\prime }}+1)}^{-\frac
12(\kappa +\kappa ^{^{\prime }}+1)}\left( \cosh (2r)\right),
\end{equation}
which can be written also as
\begin{equation}
c_n(r)=\frac 1{n!}\left( \cosh (r)\right) ^{-(\kappa +\kappa
^{^{\prime }}+1)}\left( \frac{\tanh r}r\right) ^n.
\end{equation}
The coherent states of Perelomov's type take the form
\begin{eqnarray}
\left| z,\alpha \right\rangle  &=&\left( 1-\tanh ^2\left| z\right|
\right) ^{\frac 12(\kappa +\kappa ^{^{\prime
}}+1)}\sum_{n=0}^{+\infty }\left( \frac{z\tanh \left| z\right|
}{\left| z\right| }\right) ^n\times   \nonumber
\\
&&\ \left[ \frac{\Gamma \left( n+\kappa +\kappa ^{^{\prime }}+1\right) }{%
\Gamma \left( n+1\right) \Gamma \left( \kappa +\kappa ^{^{\prime
}}+1\right) }\right] ^{\frac 12}e^{-i\alpha n(n+\kappa +\kappa
^{^{\prime }})}\left| \psi _n\right\rangle.
\end{eqnarray}
Finally, setting $\mathbf{\zeta }=\frac{z\tanh \left| z\right| }{\left|
z\right| }$, we obtain
\begin{eqnarray}
\left| \mathbf{\zeta },\alpha \right\rangle =\left( 1-\left|
\mathbf{\zeta }\right| ^2\right) ^{\frac 12(\kappa +\kappa
^{^{\prime
}}+1)} \sum\limits_{n=0}^{+\infty }\mathbf{\zeta }^n\left[ \frac{%
\Gamma \left( n+\kappa +\kappa ^{^{\prime }}+1\right) }{\Gamma
\left( n+1\right) \Gamma \left( \kappa +\kappa ^{^{\prime
}}+1\right) }\right] ^{\frac 12}e^{-i\alpha n(n+\kappa +\kappa
^{^{\prime }})}\left| \psi _n\right\rangle
\end{eqnarray}
We note that the parameter $\mathbf{\zeta }$ belongs to the unit disk $%
D=\left\{ \mathbf{\zeta \in C},\hbox{ }\left| \mathbf{\zeta
}\right| <1\right\} .$

The states are stable temporally. Indeed
\begin{equation}
e^{-iHt}\left| \mathbf{\zeta },\alpha \right\rangle =\left| \mathbf{\zeta }%
,\alpha +t\right\rangle.
\end{equation}
The identity resolution is given by
\begin{equation}
\int \left| \mathbf{\zeta },\alpha \right\rangle \left\langle
\mathbf{\zeta },\alpha \right| d\mu \left( \mathbf{\zeta }\right)
=I_{\mathcal{H}},
\end{equation}
where the measure is
\begin{equation}
d\mu \left( \mathbf{\zeta }\right) =\frac{\kappa +\kappa ^{^{\prime }}}\pi
\frac{d^2\mathbf{\zeta }}{\left( 1-\left| \mathbf{\zeta }\right| ^2\right) ^2%
}.
\end{equation}
There are two main consequence arising from the former result. First, we can
express any coherent state $\mid \mathbf{\zeta }^{^{\prime }},\alpha
^{^{\prime }}\rangle $ in terms of the others
\begin{equation}
\mid \mathbf{\zeta }^{^{\prime }},\alpha ^{^{\prime }}\rangle
=\int \left| \mathbf{\zeta },\alpha \right\rangle \left\langle
\mathbf{\zeta },\alpha \right| \mathbf{\zeta }^{^{\prime }},\alpha
^{^{\prime }}\rangle d\mu \left( \mathbf{\zeta }\right).
\end{equation}
The kernel $\left\langle \mathbf{\zeta },\alpha \right| \mathbf{\zeta }%
^{^{\prime }},\alpha ^{^{\prime }}\rangle $ is easy to evaluate from $\left(
127\right) $%
\begin{eqnarray}
\left\langle \mathbf{\zeta },\alpha \right| \mathbf{\zeta }^{^{\prime
}},\alpha ^{^{\prime }}\rangle  &=&\sqrt{\left( 1-\left| \mathbf{\zeta }%
\right| ^2\right) ^{(\kappa +\kappa ^{^{\prime }}+1)}\left( 1-\left| \mathbf{%
\zeta }^{^{\prime }}\right| ^2\right) ^{(\kappa +\kappa ^{^{\prime }}+1)}}%
\sum\limits_{n=0}^{+\infty }\overline{\mathbf{\zeta }}^n\mathbf{\zeta }%
^{^{\prime n}}\times   \nonumber \\
&&\frac{\Gamma \left( n+\kappa +\kappa ^{^{\prime }}+1\right) }{\Gamma
\left( n+1\right) \Gamma \left( \kappa +\kappa ^{^{\prime }}++1\right) }%
e^{-i(\alpha ^{^{\prime }}-\alpha )n(n+\kappa +\kappa ^{^{\prime
}})}.
\end{eqnarray}
The coherent states are normalized $\left( \left\langle \mathbf{\zeta }%
,\alpha \right| \mathbf{\zeta },\alpha \rangle =1\right) $, but they are not
orthogonal to each other.

Second, an arbitrary element state of the Hilbert space $\mathcal{H}$ , let
us call it $\left| f\right\rangle $, can be written in terms of the coherent
states
\begin{equation}
\left| f\right\rangle =\int \left( 1-\left| \mathbf{\zeta }\right|
^2\right) ^{\frac 12(\kappa +\kappa ^{^{\prime }}+1)}f(\overline{\mathbf{%
\zeta }},\alpha )\left| \mathbf{\zeta },\alpha \right\rangle d\mu
\left( \mathbf{\zeta }\right),
\end{equation}
where the analytic function
\begin{eqnarray}
f\left( \overline{\mathbf{\zeta }},\alpha \right) &=&\left( 1-\left| \mathbf{%
\zeta }\right| ^2\right) ^{-\frac 12(\kappa +\kappa ^{^{\prime
}}+1)}\left\langle \overline{\mathbf{\zeta }},\alpha \right|
f\rangle \nonumber \\ &=&\sum\limits_{n=0}^{+\infty }\mathbf{\zeta
}^n\left[ \frac{\Gamma \left( n+\kappa +\kappa ^{^{\prime
}}+1\right) }{\Gamma \left( n+1\right) \Gamma \left( \kappa
+\kappa ^{^{\prime }}+1\right) }\right] ^{\frac 12}e^{i\alpha
n(n+\kappa +\kappa ^{^{\prime }})}\left\langle \psi _n\right|
f\rangle
\end{eqnarray}
determines in a complete way the state $\left| f\right\rangle \in $ $\mathcal{%
H}$. The state $\left| \psi _n\right\rangle $ is represented by the function

\begin{equation}
\mathcal{F}_n^{\prime }\left( \mathbf{\zeta },\alpha \right) =\mathbf{\zeta }%
^n\left[ \frac{\Gamma \left( n+\kappa +\kappa ^{^{\prime }}+1\right) }{%
\Gamma \left( n+1\right) \Gamma \left( \kappa +\kappa ^{^{\prime
}}+1\right) }\right] ^{\frac 12}e^{i\alpha n(n+\kappa +\kappa
^{^{\prime }})}.
\end{equation}
The creation $a_{\kappa ,\kappa ^{^{\prime }}}^{+}$ annihilation
$a_{\kappa ,\kappa ^{^{\prime }}}^{-}$ and $G_{\kappa ,\kappa
^{^{\prime }}}\left( N\right) $ operators act in the Hilbert space
of analytic functions $f\left( \mathbf{\zeta },\alpha \right) $ as
a first-order differential operators:

\begin{eqnarray}
a_{\kappa ,\kappa ^{^{\prime }}}^{+} &=&\mathbf{\zeta }^2\frac d{d\mathbf{%
\zeta }}+(\kappa +\kappa ^{^{\prime }}+1)\mathbf{\zeta },%
\hspace{1cm}a_{\kappa ,\kappa ^{^{\prime }}}^{-}=\frac
d{d\mathbf{\zeta }} \nonumber, \\
G_{\kappa ,\kappa ^{^{\prime }}}\left( N\right) &\equiv &G=2\mathbf{\zeta }%
\frac d{d\mathbf{\zeta }}+(\kappa +\kappa ^{^{\prime }}+1).
\end{eqnarray}
One can verify that
\begin{eqnarray}
a_{\kappa ,\kappa ^{^{\prime }}}^{+}\mathcal{F}_n^{\prime }\left( \mathbf{%
\zeta },\alpha \right) &=&\sqrt{\left( n+1\right) \left( n+1+\kappa +\kappa
^{^{\prime }}\right) }e^{-i\alpha (2n+1+\kappa +\kappa ^{^{\prime }})}%
\mathcal{F}_{n+1}^{\prime }\left( \mathbf{\zeta },\alpha \right) \\
a_{\kappa ,\kappa ^{^{\prime }}}^{-}\mathcal{F}_n^{\prime }\left( \mathbf{%
\zeta },\alpha \right) &=&\sqrt{n\left( n+\kappa +\kappa ^{^{\prime
}}\right) }e^{i\alpha (2n-1+\kappa +\kappa ^{^{\prime }})}\mathcal{F}%
_{n-1}^{\prime }\left( \mathbf{\zeta },\alpha \right), \\
G_{\kappa ,\kappa ^{^{\prime }}}\left( N\right)
\mathcal{F}_n^{\prime
}\left( \mathbf{\zeta },\alpha \right) &=&(2n+1+\kappa +\kappa ^{^{\prime }})%
\mathcal{F}_n^{\prime }\left( \mathbf{\zeta },\alpha \right).
\end{eqnarray}
The analytic representation of the Gazeau-Klauder coherent states and the
analytical realization of the Perelomov ones in the unit disk are related
through a Laplace transform. Indeed one can verify easily that

\begin{equation}
\mathcal{F}_n^{\prime }\left( \mathbf{\zeta },\alpha \right) =\frac{\zeta
^{-(\kappa +\kappa ^{\prime }+1)}}{\sqrt{\Gamma (\kappa +\kappa ^{\prime }+1)%
}}\int_0^{+\infty }z^{\kappa +\kappa ^{\prime
}}\mathcal{F}_n\left( z,\alpha \right) e^{-\frac z\zeta }dz,
\end{equation}
which means that the function $\mathcal{F}_n^{\prime }\left( \frac 1{\mathbf{%
\zeta }},\alpha \right) $ is the Laplace transform of $z^{\kappa +\kappa
^{\prime }}\mathcal{F}_n\left( z,\alpha \right) $. A similar result was
obtained in $\left[ 28\right] $ showing that the representation in the unit
disk and Barut-Girardello one, based on the $su(1,1)$ coherent states, are
related through a Laplace transform.

\subsection{P\"oschl-Teller generalized intelligent states}

The generalized intelligent states can be determined by using two
analytic representation, one based on the so-called Gazeau-Klauder
coherent states (section 3) and the other one on the Perelomov's
coherent states (section 4).

\subsubsection{The Gazeau-Klauder analytic representation}

We introduce the analytic function
\begin{equation}
\Phi _{\left( z^{\prime },\lambda ,\alpha \right) }\left( z\right) =\sqrt{%
\frac{I_{\kappa +\kappa ^{^{\prime }}}\left( 2\left| z\right| \right) }{%
\left| z\right| ^{\kappa +\kappa ^{\prime }}}}\langle \overline{z},\alpha
\left| z^{\prime },\lambda ,\alpha \right\rangle
\end{equation}
by mean of which one convert the eigenvalues equation
\begin{equation}
\left[ \left( 1+\lambda \right) a_{\kappa ,\kappa ^{\prime }}^{-}+\left(
1-\lambda \right) a_{\kappa ,\kappa ^{\prime }}^{+}\right] \left| z^{\prime
},\lambda ,\alpha \right\rangle =2z^{\prime }\left| z^{\prime },\lambda
,\alpha \right\rangle
\end{equation}
into the second-order linear homogeneous differential equation
\begin{equation}
\left[ \left( 1+\lambda \right) \left( z\frac{d^2}{dz^2}+\left(
\kappa +\kappa ^{\prime }+1\right) \frac d{dz}\right) +\left(
1-\lambda \right) z-2z^{\prime }\right] \Phi _{(z^{\prime
},\lambda )}\left( z\right) =0.
\end{equation}
We first consider the general case $\lambda \neq \pm 1$. Setting

\begin{equation}
\Phi _{(z^{\prime },\lambda )}\left( z\right) =\exp \left( \pm \sqrt{\frac{%
\lambda -1}{\lambda +1}}z\right) F_{(z^{\prime },\lambda )}(z),
\end{equation}
The equation can be transformed in the Kummer equation
\begin{equation}
\left[ Z\frac{d^2}{dZ^2}+\left( \kappa +\kappa ^{\prime }+1-Z\right) \frac
d{dZ}-\left( \frac{\kappa +\kappa ^{\prime }+1}2\mp \frac{z^{\prime }}{%
\sqrt{\lambda ^2-1}}\right) \right] F_{(z^{\prime },\lambda
)}\left( z\right) =0,
\end{equation}
where $Z=\mp 2\sqrt{\frac{\lambda -1}{\lambda +1}}z$

Then the solutions of the equation $\left( 143\right) $ are given by
\begin{equation}
\Phi _{(z^{\prime },\lambda )}\left( z\right) =\exp \left( \pm \sqrt{\frac{%
\lambda -1}{\lambda +1}}z\right) \hbox{ }_1F_1\left( \frac{\kappa
+\kappa ^{\prime }+1}2\mp \frac{z^{\prime }}{\sqrt{\lambda
^2-1}},\kappa +\kappa ^{\prime }+1;\mp 2\sqrt{\frac{\lambda
-1}{\lambda +1}}z\right)
\end{equation}
or
\begin{equation}
\Phi _{(z^{\prime },\lambda )}\left( z\right) =\exp \left( \pm \sqrt{\frac{%
\lambda -1}{\lambda +1}}z\right) z^{-(\kappa +\kappa^{\prime})}\hbox{ }
_1F_1\left( \frac{1-(\kappa +\kappa ^{\prime })}2\mp \frac{z^{\prime }}{%
\sqrt{\lambda ^2-1}},1-(\kappa +\kappa ^{\prime });\mp
2\sqrt{\frac{\lambda -1}{\lambda +1}}z\right)
\end{equation}
The first solution $\left( 146\right) $ is always analytic, but the solution
$\left( 147\right) $ is not (Remember that $\kappa >1$ and $\kappa ^{\prime
}>1$). The upper and lower signs in equation $\left( 146\right) $ are
equivalent, because the confluent hypergeometric function $_1F_1(\alpha
,\gamma ;z)$ can be written in two equivalents forms which are related by
Kummer's transformation
\begin{equation}
_1F_1(\alpha ,\gamma ;z)=e^z\hbox{ }_1F_1(\gamma -\alpha ,\gamma
,-z).
\end{equation}
Using the properties of this hypergeometric functions, we conclude that the
squeezing parameter $\lambda $ obeys to the condition
\begin{equation}
\sqrt{\left| \frac{\lambda -1}{\lambda +1}\right| }<1\Leftrightarrow %
\hspace{1cm}\hbox{Re}(\lambda )>0,
\end{equation}
which exactly the restriction on $\lambda $ imposed by the positivity of the
commutator $\left[ a_{\kappa ,\kappa ^{^{\prime }}}^{-},a_{\kappa ,\kappa
^{^{\prime }}}^{+}\right] =G_{\kappa ,\kappa ^{^{\prime }}}\left( N\right) $
(see equations $\left( 109\right) $ and $\left( \mathrm{110}\right) $).

We consider now the degenerate cases $\lambda =-1$ and $\lambda =1.$ For the
$\lambda =-1$ the equation $\left( \mathrm{143}\right) $ does not have any
normalized analytic solution (the operator $a_{\kappa ,\kappa ^{\prime }}^{+}
$ does not have any eigenstate). For $\lambda =1,$ using the power series of
$_1F_1(a,b;z)$, we get
\begin{equation}
\Phi _{(z^{\prime },\lambda =1)}\left( z\right) =\hbox{
}_0F_1(\kappa +\kappa ^{\prime }+1;zz^{\prime }).
\end{equation}
The result $\left( 150\right) $ coincides with the solution $\left(
111\right) $ (up to normalization constant) for $\lambda =1$, and we recover
the P\"oschl-Teller coherent states defined as the $a_{\kappa ,\kappa
^{\prime }}^{-}$ eigenstates.

\subsubsection{The Perelomov coherent state basis and analytic
representation in the unit disk}

To solve the eigenvalues equation $\left( \mathrm{142}\right) $, using the
analytic representation of Perelomov coherent states in the unit disk, we
introduce the analytic function
\begin{equation}
\Phi _{(\mathbf{\zeta }^{\prime },\lambda )}\left( \mathbf{\zeta }\right)
=\left( 1-\left| \mathbf{\zeta }\right| ^2\right) ^{-\frac 12(\kappa +\kappa
^{^{\prime }}+1)}\langle \overline{\mathbf{\zeta }},\alpha \left| \mathbf{%
\zeta }^{\prime },\lambda ,\alpha \right\rangle.
\end{equation}
Equation $\left( 142\right) $ is then converted to the following
differential equation

\begin{equation}
\left[ \left[ (1-\lambda )\mathbf{\zeta }^2+(1+\lambda )\right] \frac d{d%
\mathbf{\zeta }}+(1-\lambda )(\kappa ^{\prime }+\kappa +1)\mathbf{\zeta }-2%
\mathbf{\zeta }^{\prime }\right] \Phi _{(\mathbf{\zeta }^{\prime
},\lambda )}(\mathbf{\zeta })=0.
\end{equation}
Admissible values of $\lambda $ and $\mathbf{\zeta }^{\prime }$ are
determined by the requirements that the functions $\Phi _{(\mathbf{\zeta }%
^{\prime },\lambda )}(\mathbf{\zeta })$ must be analytic in the unit disk.
We consider the general case. The solution of Eq. $\left( \mathrm{152}%
\right) $ is

\begin{equation}
\Phi _{(\mathbf{\zeta }^{\prime },\lambda )}(\mathbf{\zeta })=\mathcal{N}%
^{-\frac 12}\prod_{l=\pm 1}\left( 1+l\left( \frac{\lambda -1}{\lambda +1}%
\right) ^{\frac 12}\mathbf{\zeta }\right) ^{-\frac 12(\kappa
+\kappa ^{\prime }+1)+l\frac{\mathbf{\zeta }^{\prime
}}{\sqrt{\lambda ^2-1}}},
\end{equation}
where $\mathcal{N}$ is a normalization constant. The condition of
analyticity requires

\begin{equation}
\left| \frac{\lambda -1}{\lambda +1}\right| <1\Leftrightarrow \hbox{ Re}%
\lambda >0.
\end{equation}
If Re$\lambda <0$, the function $\Phi _{(\mathbf{\zeta }^{\prime },\lambda
)}(\mathbf{\zeta })$ cannot be analytic in the unit disk.

The decomposition of the generalized intelligent states $\left| \mathbf{%
\zeta }^{\prime },\lambda ,\alpha \right\rangle $ over the Hilbert
orthonormal basis \{$\left| \psi _n\right\rangle $\} can be obtained by
expanding the function $\Phi _{(\mathbf{\zeta }^{\prime },\alpha )}(\mathbf{%
\zeta })$ into a power series in $\mathbf{\zeta }.$ This can be done using
the following relations

\begin{equation}
\left( 1+\left( \frac{\lambda -1}{\lambda +1}\right) ^{\frac 12}\mathbf{%
\zeta }\right) ^{\alpha _{+}}\left( 1-\left( \frac{\lambda -1}{\lambda +1}%
\right) ^{\frac 12}\mathbf{\zeta }\right) ^{\alpha
_{-}}=\sum\limits_{n=0}^{+\infty }\mathbf{\zeta }^n\left( 2\sqrt{\frac{%
\lambda -1}{\lambda +1}}\right) ^nP_n^{(\alpha _{+}-n,\alpha
_{-}-n)}(0),
\end{equation}
where
\begin{equation}
\alpha _{\pm }=-\frac 12(\kappa +\kappa ^{\prime }+1)\pm
\frac{\mathbf{\zeta }^{\prime }}{\sqrt{\lambda ^2-1}}.
\end{equation}
Then, the function $\Phi _{(\mathbf{\zeta }^{\prime },\alpha )}(\mathbf{%
\zeta })$ can be expanded in terms of the Jacobi polynomials $P_n^{(\alpha
,\beta )}(x)$. Using the relation between the hypergeometric function and
Jacobi polynomials $\left[ \mathrm{27}\right] $, one can show

\begin{eqnarray}
\left| \mathbf{\zeta }^{\prime },\lambda ,\alpha \right\rangle  &=&\mathcal{N%
}^{-\frac 12}\sum\limits_{n=0}^{+\infty }\left[ \frac{\Gamma
(\kappa +\kappa ^{\prime }+1)}{n!\Gamma (\kappa +\kappa ^{\prime
}+1+n)}\right]
^{\frac 12}\left[ \frac{\Gamma (\alpha _{+}+1)}{\Gamma (\alpha _{+}-n+1)}%
\right] \left( 2\sqrt{\frac{\lambda -1}{\lambda +1}}\right) ^n\times
\nonumber \\
&&_2F_1(-n,-n-(\kappa +\kappa ^{\prime }),\alpha _{+}-n+1;\frac
12)e^{-i\alpha E_n}\left| \psi _n\right\rangle
\end{eqnarray}
or
\begin{equation}
\left| \zeta ^{\prime },\lambda ,\alpha \right\rangle
=\mathcal{N}^{-\frac 12}\sum\limits_{n=0}^{+\infty }\left[
\frac{n!\Gamma (\kappa +\kappa ^{\prime }+1)}{\Gamma (\kappa
+\kappa ^{\prime }+1+n)}\right] ^{\frac 12}\left(
2\sqrt{\frac{\lambda -1}{\lambda +1}}\right) ^nP_n^{(\alpha
_{+}-n,\alpha _{-}-n)}(0)e^{-i\alpha E_n}\left| \psi
_n\right\rangle.
\end{equation}
The generalized intelligent states $\Phi _{(\mathbf{\zeta }^{\prime
},\lambda )}(\mathbf{\zeta )}$ and $\Phi _{(z^{\prime },\lambda )}(z\mathbf{)%
}$ are related through a Laplace transform. In fact, equation $\left(
\mathrm{152}\right) $ can be written as

\begin{equation}
\left[ \left[ (1+\lambda )\mathbf{\zeta }^2+(1-\lambda )\right] \frac d{d%
\mathbf{\zeta }}-\frac{(1-\lambda )(\kappa ^{\prime }+\kappa +1)}{\mathbf{%
\zeta }}+2\mathbf{\zeta }^{\prime }\right] \Phi _{(\mathbf{\zeta
}^{\prime },\lambda )}\left( \frac 1{\mathbf{\zeta }}\right) =0.
\end{equation}
Using
\begin{equation}
\Phi _{(\mathbf{\zeta }^{\prime },\lambda )}\left( \frac 1{\mathbf{\zeta }%
}\right) =\frac{\mathbf{\zeta }^{-(\kappa +\kappa ^{\prime }+1)}}{\sqrt{%
\Gamma (\kappa +\kappa ^{\prime }+1)}}\int_0^{+\infty }z^{\kappa
+\kappa ^{\prime }}\Phi _{(\mathbf{\zeta }^{\prime },\lambda
)}(z\mathbf{)}e^{-\frac z{\mathbf{\zeta }}}dz.
\end{equation}
It is easy to see that the eigenvalues equation $\left(
\mathrm{159}\right) $ becomes

\begin{equation}
\left[ \left( 1+\lambda \right) \left( z\frac{d^2}{dz^2}+\left( \kappa
+\kappa ^{\prime }+1\right) \frac d{dz}\right) +\left( 1-\lambda \right) z-2%
\mathbf{\zeta }^{\prime }\right] \Phi _{(\mathbf{\zeta }^{\prime
},\lambda )}\left( z\right) =0,
\end{equation}
which coincides with $\left( 143\right) $ ones ($\mathbf{\zeta }^{\prime
}=z^{\prime }$) that gives the generalized intelligent states $\left(
146\right) .$

\section{Summary}

In this work, we have explicitly constructed the Gazeau-Klauder and
Perelomov coherent states for an arbitrary quantum system. As an
application, of this construction, we considered the system trapped in the
P\"oschl-Teller potentials type. We shown that the analytical
representations of Gazeau-Klauder and Perelomov coherent states (which are
related through a Laplace transform) enables us to compute the generalized
intelligent states for the P\"oschl-Teller potentials. Finally, it should be
interesting to investigate further applications of the results obtained on
this work. Indeed, it is interest, in our opinion, to construct the coherent
states and generalized intelligent states for the Shape invariant potentials
$\left[ 29\right] $. This matter will be considered in a forthcoming work.

\begin{quote}
\textbf{Acknowledgments}

The authors are grateful to the referee for critical comments and extensive
suggestions on an earlier draft of the manuscript, which helped greatly to
improve the clarity of the presentation. The senior author M. Daoud would
like to thank V. Hussin for valuable discussions.
\end{quote}

\newpage\

\end{document}